\documentclass[a4paper,fleqn,usenatbib]{mnras}
\usepackage{graphicx}
\usepackage{amsmath}
\usepackage{amssymb}
\usepackage[T1]{fontenc}
\usepackage{ae,aecompl}
\usepackage{graphicx}
\usepackage{newtxtext,newtxmath}

\begin{document}

\title[What goes around...]{Consequences of dynamically unstable moons in extrasolar systems}
\author[Hansen]{
        Bradley M. S. Hansen,$^{1}$\thanks{E-mail:hansen@astro.ucla.edu},
        \\
        $^{1}$Mani L. Bhaumik Institute for Theoretical Physics, Department of Physics and Astronomy, University of California, Los Angeles, CA, 90095, USA\\
}

\date{Accepted XXX. Received YYY; in original form ZZZ}

\maketitle

\begin{abstract}
Moons orbiting rocky exoplanets in compact orbits about other stars experience an accelerated tidal evolution, and can either merge with their parent planet or reach the limit
of dynamical instability within a Hubble time.
 We review the parameter space over which moons become unbound, including the effects of atmospheric tides on the planetary spin. We find that such tides can change the final outcome from merger to escape, albeit over a limited parameter space. We also follow the further evolution of unbound moons, and
 demonstrate that the overwhelmingly most likely long-term outcome is that the unbound
moon returns to collide with its original parent planet. The dust released by such a collision is estimated to reach optical depths $\sim 10^{-3}$, exhibit characteristic temperatures
of a few hundred degrees Kelvin, and last for a few thousand years. These properties make such events an attractive model for the emerging class of middle-aged main sequence stars that are observed to 
show transient clouds of warm dust. Furthermore, a late  collision between a planet and a returning moon on a hyperbolic orbit may  sterilise an otherwise habitable planet.
\end{abstract}


\section{Introduction}

The planets of the Solar System are accompanied by a variety of satellites -- from the moons and rings of the giant planets to the compact
moons of Earth and  Mars \citep{Peale99}. The fact that such structures are relatively common in the Solar system suggests that at least some fraction of
the planets formed around other stars must be, in turn, accompanied by their own moons and rings. These satellites also offer insights into the 
processes that sculpt planetary systems -- a subject of great interest at present as observations push our knowledge well past the parameter
space occupied by the Solar system planets.
 Although of great theoretical interest, the low mass ratios and small sizes imply that
most exoplanetary moon systems are likely to remain unobservable for some time. Studies of transitting giant planets \citep{SSS07,Kip09,AK13,Lew13,HWK14,Hipp15,AJ15} are now beginning to probe
the upper plausible mass range of exomoons \citep{TK18} although it is not clear whether such masses can be explained by
traditional processes \citep{ONI14,Ham18,H19,Cili21}.

The moons of rocky planets are of particular interest. The formation of compact, low-mass, planetary systems is still a question of active discussion. If the late
stages of planetary assembly occur in situ \citep{HM12,HM13,CL13}, then moon formation is expected to be a natural consequence of the giant impacts that occur
during the final stages of planetary system clearing. On the other hand, if the planets migrate a significant distance inwards to their current configurations
\citep{IL10,IOR17}, then the additional torques experienced by the system may render systems unstable \citep{SBA16}. In this sense, the presence of moons in compact, extrasolar planetary systems may serve as a marker of a late stage of in situ giant impacts.
The moons of terrestrial planets may also play a role in preserving planetary habitability. The large lever arm of a moon orbit is 
 potentially a stabilising force against chaotic variations in planetary obliquity caused by resonance between planetary precession frequencies and the secular eigenmodes of planetary systems \citep{LJR93}. However, at least in the case of the Earth,  it may not be absolutely necessary \citep{LBC12}.

An important consideration in extending the study of satellites to other, compact, planetary systems is that the tidal evolution of moon systems proceeds at a faster pace for planets that are close to their
parent stars \citep{BOB02,SBO12}, although this depends somewhat on the parameters of the tidal model \citep{Piro18,TP20}.
At some level this question is also moot, because terrestrial class planets are themselves difficult to detect, and so the detection of the moons
of extrasolar terrestrial in place is still somewhat beyond current capabilities. However, if these systems are short lived, there is the possibility
that we may observe the destruction of such objects, and thereby understand something about their frequency of occurrence.

In particular, moons that spiral out to the point at which the circumplanetary orbit becomes unstable will enter an orbit around the star with approximately 
the same semi-major axis as the parent planet.
Although no longer bound to the original planet, they will still be bound to the host star. As such, they will therefore undergo repeated close passages and scatter off the planet.  For a moon freed
from a massive planet on scales of several AU, this may result in the ejection of the moon from the system. However, for 
the kinds of compact systems susceptible to tidal evolution, the potential well of the star is too deep for planetary scattering for ejection
to be a significant loss channel. As such, 
 nearly all of the freed exomoons will eventually collide with the original host planet and we anticipate a dramatic release of dust would result, and the formation of a temporary
infra-red excess around the parent star. Such a scenario may offer an explanation for the class of objects known as EDD \citep{Balog09,Melis16,Kral17,Moor21,MOS21} hereafter, which are
middle aged stars (100 Myr or older) that show  infrared excesses well in excess of that expected from the traditional models of planetary
system assembly. It is the quantitative aspects of this scenario that we wish to investigate here.

In \S~\ref{Unstable} we review the evolution of plausibe moons in  extrasolar rocky planet systems, and examine the expected age range over which
they go unstable. In addition to the traditional stellar tides discussed by previous authors \citep{BOB02,SBO12,TP20}, we also consider the effects of atmospheric tides, which have been
postulated \citep{GS69,CL01,Leconte15,CCL15,ALMC17} to provide asynchronous equilibria  around lower mass stellar hosts. In \S~\ref{Orbits} we discuss the post-ejection
evolution of unstable moons, and consider their ultimate fates.
 In \S~\ref{EDD} we discuss the potential observational signatures of such fates.

\section{Orbital Evolution of Moons}
\label{Unstable}

The basic picture for a terrestrial-type planet+moon system is that the moon forms as the result of a collision during the late stages of planet
assembly. A vapor disk produced in the collision condenses into solid particles at the appropriate Roche radius and these bodies rapidly co-alesce
into a moon \citep{DS89,Canup21}. The collision also leaves the planet with significant spin and the long-term evolution of the moon is determined by the tidal coupling
between the planet and the moon, with three limiting outcomes available \citep{Couns73}. If the planet spin is too low, the planet will extract angular
momentum from the orbit and the moon will spiral in to merge with the planet. If the planet spin is high enough, angular momentum will be transferred
to the moon and it will spiral outwards, eventually to achieve a synchronous rotation or to spiral outwards towards escape. Within the context of a
planet orbiting a star, this outwards evolution is truncated when the separation approaches the Hill sphere and the orbit becomes dynamically unstable.

We now know of many planetary systems where the planets are much closer to the host star than the planets in our own Solar system.
In such cases, an additional influence must be accounted for. The host star also raises a tide on the planet,
and so dissipation of the planetary tide also acts to transfer angular momentum from the spin of the planet to its orbit around the star.
This acts to spin the planet down, and so can cause the outwards orbital evolution to reverse, increasing the range of parameter space
leading to inspiral. In this scenario, achieving a synchronous state is also unlikely, because synchronism between the planetary spin, planetary
orbital frequency and moon orbital frequency would place the synchronous orbit at the approximate location of the Hill sphere. 


\subsection{Orbital Evolution Equations}

There have been several prior calculations of the orbital evolution of moons due to tides \citep{BOB02,SBO12,Piro18,TP20}, and they all give the same qualitative behaviour, but can
differ in quantitative ways, primarily because of variations in the assumption of
tidal strengths. As in prior works, the orbital evolution of the moon is determined by the gravitational torque acting to transfer angular momentum
between the satellite orbit and the planetary spin
\begin{equation}
\frac{d}{dt} \left( M_m (G M_p a_m)^{1/2} +  I_m n_m\right) = - N_m
\end{equation}
where $M_m$, $a_m$, $I_m$ and $n_m$ represent the moon mass, semi-major axis, moment of inertia and orbital frequency respectively. The torque $N_m$
is driven by dissipation in the planet and is given by
\begin{equation}
N_m = \frac{3 k_2}{2 Q} \frac{G M_m^2 R_p^2}{a_m^6} b_m(n_m,\Omega_p)
\end{equation}
where $k_2$ and $Q$ are the tidal Love number and quality factor respectively. The function $b_m$ encodes the frequency dependance of the tidal response
and will be discussed below. 

The orbital evolution of the moon is coupled to the spin evolution of the planet, and so $N_m$ will also act on the spin of the planet. In addition, planets that orbit close to their host stars will also experience a gravitational tide due to the influence of the host star, leading to the operation of a 
second torque (also driven by dissipation in the planet)
\begin{equation}
N_p =  \frac{3 k_2}{2 Q} \frac{G M_*^2 R_p^2}{a_p^6} b_p(n_p,\Omega_p)
\end{equation}
where $M_*$ is the mass of the host star and $R_p$, $a_p$, $n_p$ are the radius, semi-major axis and orbital frequency for the planet. Similarly, the
function $b_p$ encodes the frequency response. 

In addition to the gravitational tide, planets with thick atmospheres may experience an atmospheric tide, driven by the heating of the atmosphere by stellar irradiation, and regulated by the forces exerted by the atmosphere on the solid body rotation of the planet. This torque has long been held responsible for the peculiar retrograde spin of Venus in our solar system \citep{ID78,DI80,CL01} and may also apply more generally for planets in close orbits \citep{Leconte15,CCL15,ALMC17}. We therefore include an atmospheric torque on the planetary spin
\begin{equation}
N_a = -\frac{3}{2} q_a \frac{3 M_* R_p^3}{5 \bar{\rho} a_p^3} b_a(n_p,\Omega_p) \label{eqn:AtmoT}
\end{equation}
where $\bar{\rho}$ is the planetary mean density, $q_a$ is the amplitude of pressure field at the base of the planetary atmosphere, and $b_a$ represents
the frequency dependance.  The sum of the torques $N_m$, $N_p$ and $N_a$ will determine the evolution of the planetary spin, which will, in turn, drive the orbital evolution of the moon. 

One important consideration is the amplitude and frequency dependance of the various torques, which is determined by the physical mechanisms that underlie the dissipation in the planet. For an Earth-like planet, the dissipation is believed to result from the dissipation in the boundary layers of marginal seas \citep{Jeff21} and resulting from pelagic turbulence \citep{Bell75,ER00}, with a relatively weak frequency response. These dissipation mechanisms require the existence of oceans and dry planets may have weaker dissipation, as estimated for Mars from \cite{LDP07}. On the other hand, the presence of substantial Hydrogen envelopes observed on many extrasolar planets may enable the propagation of gravity waves, causing dissipation from the conversion of baroclinic waves into barotropic waves on topographic features \citep{Bell75,BIY02}. In the light of the various potential mechanisms, we will adopt a broad parameterisation, calibrated in terms of the tidal Q parameter ($Q_{\oplus}=30$ serving as a benchmark). We will assume $Q$ is frequency independent, but changes sign with $n_p-\Omega_p$ or $n_m-\Omega_p$. Many previous treatments assume that this occurs as a step function at zero frequency. However, studies of the frequency response of solid bodies \citep{Efro12} suggest that, at low enough frequencies, the tidal response is driven by the internal viscosity of the body. As such, we will employ a smoother transition (an arctangent function), parameterised by the viscous timescale of Earth-like planets. 
For the atmospheric tide model, we adopt the results of \cite{Leconte15}, which are determined from global circulation models under different levels of irradiation and atmospheric pressures. Our default model is for an atmospheric surface pressure of 1 bar, and a level of irradiation appropriate to the inner habitable zone.  Appendix~\ref{TidalModel} describes the atmospheric torque in quantitative terms and demonstrates the kinds of spin equilibria that can result.

In normalised form, our equation for the evolution of the moon orbit is
\begin{equation}
\frac{dx}{d\tau}  =  \frac{1}{x^7} \left( y x^{3/2} - 1 \right) \bar{b}_m \label{xevol} 
\end{equation}
 where $x=a_m/R_{\oplus}$, $y=\Omega_p/\Omega_k$, 
 $\tau=t/t_0$ and 
\begin{eqnarray}
\Omega_k&=&\left( \frac{G M_{\oplus}}{R_{\oplus}^3} \right) \\
t_0&=&0.068 {\rm yrs} \frac{Q}{30}  \left( \frac{M_m/M_p}{0.0123} \right)^{-1} \left( \frac{M_p}{M_{\oplus}} \right)^{-1/2} 
\left( \frac{R_p}{R_{\oplus}} \right)^{-5}
\end{eqnarray}
with $\alpha_p$ is the moment of inertia constant for the planet. The function $\bar{b}_m$ is
\begin{equation}
\bar{b}_m = \frac{2}{\pi} \arctan\left( \gamma_0 (x^{-3/2}-y)\right) \label{barbm}
\end{equation}
where $\gamma_0=10\Omega_k \tau_M = 10^5$ for $\tau_M=10^{10}s$ (the characteristic viscous timescale -- the Maxwell time).   
This function ensures a smooth transition from positive to negative torques as the forcing frequency drops below $1/\tau_M$.

Our equation for the evolution of the stellar spin is more complicated, with three different torques contributing.

\begin{eqnarray}
\frac{dy}{d\tau} & = & \frac{\gamma_1}{ x^{15/2}} \left( y x^{3/2} - 1  \right) \bar{b}_m + \gamma_2 \bar{b}_p + \\ 
 &  & \gamma_3 \frac{ \left(y - y_p\right)}{1 + \gamma_4 \left( y - y_p \right)^2} \label{yevol}
\end{eqnarray}
and
\begin{eqnarray}
\gamma_1&=&\frac{1}{\alpha_p} \frac{M_m}{M_p} \left( \frac{R_p}{R_{\oplus}} \right)^{-2} \\
\gamma_2&=&\frac{1}{\alpha_p} \frac{M_*^2}{M_m M_p} \left( \frac{R_{\oplus}}{a_p} \right)^6  \left( \frac{R_p}{R_{\oplus}} \right)^{-2} \\
\gamma_3&=&\frac{4 \pi}{5} \frac{Q \bar{q}_0}{\alpha k_2 \bar{\omega}_0} \frac{M_*}{M_p} \frac{M_{\oplus}}{M_m} \frac{R_{\oplus}}{R_p} 
\left( \frac{a_p}{R_{\oplus}}\right)^{-3} \\
\gamma_4&=& 1/\bar{\omega}_0^{2} = (\Omega_k/\omega_0)^2 \\
\bar{q}_0 &=& \frac{q_0 R_{\oplus}^4}{G M_{\oplus}^2} \\
\bar{b}_p &=& \frac{2}{\pi} \arctan\left( \gamma_0 (y_p-y) \right) \label{barbp}
\end{eqnarray}
The quantities $q_0$ and $\omega_0$ are taken from \cite{Leconte15}. Moons are assumed to form at just outside the Roche radius (taken
to be $x=3$ here) and the outer limit for dynamical stability is taken to be approximately half of the Hill sphere radius
$R_H = a_p (M_p/3 M_*)^{1/3}$. This factor is based on numerous prior studies \citep{DWY06,Donn10} and confirmed with our own
calculations in \S~\ref{Orbits}.
Thus, we truncate our outward evolution because 
 dynamical instability occurs for orbits with
 or $x>x_i= 119 (a_p/1 AU) (M_p/M_{\oplus})^{1/3} (M_*/M_{\odot})^{-1/3}$.
 
\subsection{Solutions}

Our baseline calculation assumes a replica of the Earth--Moon system, orbiting the Sun at different values of the semi-major axis. The
moon evolution begins at $x=3$, with an initial spin $y=y(0)$, which then evolves according equations~(\ref{xevol}) and (\ref{yevol}).
The system is either evolved until the moon spirals in to $x=3$ again (corresponding to tidal disruption), spirals out to the dynamical
instability limit, or reaches a system age of $10^{10}$ years. 

\subsubsection{Gravitational Tides only}

If we omit the atmospheric tides, our system of equations describes a version of the calculations previously calculated \citep{BOB02,SBO12,TP20}.
The simplest evolution results in the case where the stellar tide is also weak enough to be irrelevant. In that case, the equations
have a 
 a simple analytic solution -- corresponding to basic angular momentum
conservation -- given by
\begin{equation}
 y = y(0) - 2 \gamma_1 \left( x^{1/2} - x(0)^{1/2} \right).
\end{equation}
where we assume $x(0)=3$ here.
In the limit $y x^{3/2} \gg 1$, and large $y(0)$, we can also estimate a timescale it will
take to spiral out to a particular separation $x$, namely
\begin{equation}
 \tau \sim \frac{2}{13 y(0)} x^{13/2}
\end{equation}
and so we can estimate an outspiral time 
\begin{eqnarray}
T_{esc} & = & 5.4 \times 10^{11} {\rm yrs} \left( \frac{a}{1 AU} \right)^{13/2} \left(\frac{0.0123 M_{\oplus}}{M_{m}/M_{p}} \right) \frac{Q'}{30} \nonumber \\
 && \left( \frac{M_p}{M_{\oplus}} \right)^{5/3} \left(\frac{R_p}{R_{\oplus}} \right)^{-5} \left( \frac{M_*}{M_{\odot}} \right)^{-13/6} \left( \frac{y(0)}{0.6} \right)^{-1} \label{EscapeTime}.
\end{eqnarray}
Thus, within the context of the Earth-Moon system, the moon would survive for $a > 0.48$AU (in the absence of stellar tides).

\begin{figure}
\includegraphics[height=5cm, width=5cm, angle=0, scale=1.75]{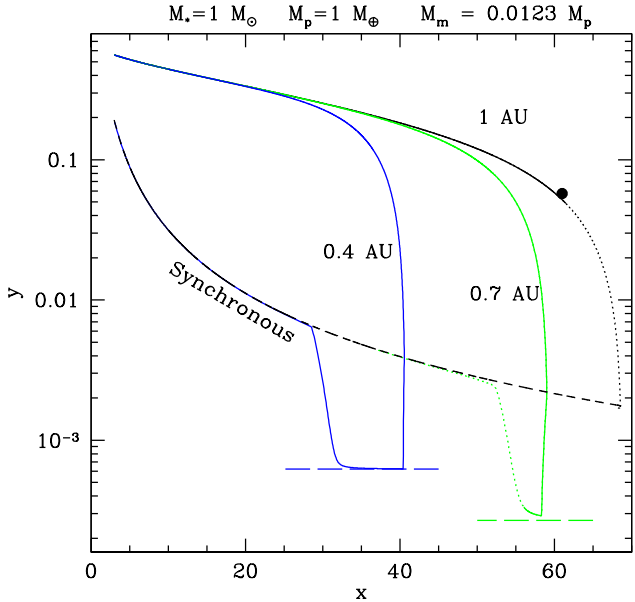}
\caption{The black, green and blue curves show the evolution of the planetary spin and moon semi-major axis for three
systems with the Earth--Moon parameters, but different heliocentric semi-major axes. The solid part of each curve extends
up to an age of 4.5~Gyr. The dotted part extends this to 10~Gyr. The short dashed line indicates the locus of spin--orbit
synchronism. The horizontal long dashed lines indicate the spin at which the planet is synchronous with its orbital frequency
around the Sun (for each case). The solid point indicates the parameters of the present Earth--Moon system.
\label{XY}}
\end{figure}

The addition of the stellar tides bleeds angular momentum from the system and causes the orbital evolution to reverse if
$y x^{3/2} < 1$.
Figure~\ref{XY} shows the nature of the solutions for the gravitational tides only case, for an Earth-like configuration
located at three different semi-major axes (1 AU, 0.7 AU and 0.4 AU). The initial spin ($y(0)=0.6$) and tidal $Q$ ($Q=30$) were chosen to
yield present-day Earth-Moon conditions (the solid point on the right). The equations have been integrated to $10^{10}$ years,
but the solid part of each curve extends only to an age of 4.5~Gyr.

We see that the Earth--Moon system has largely followed the simple solution in which angular momentum conservation is the
principal determinant of the orbital properties. At a closer distance of a=0.7~AU (approximately the distance of Venus),
we see that the stellar tides are capable of spinning the planet down to synchronism with the heliocentric orbit, and
that the moon has begun to spiral in and will eventually approach the synchronous orbit, although it will not be tidally
disrupted within 10~Gyr. This is the Type II case described by \cite{SBO12}.
 At an even closer orbit (0.4 AU) we see that the tidal evolution proceeds all the way to inspiral
and merger.

\subsubsection{Including Atmospheric Tides}

\begin{figure}
\includegraphics[height=5cm, width=5cm, angle=0, scale=1.75]{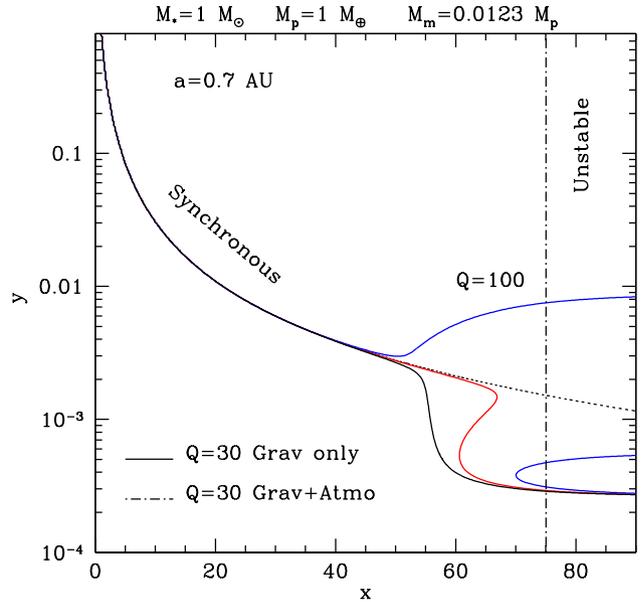}
\caption{The dotted curve shows the locus for which the sign of the orbital evolution reverses, i.e. $dx/dt=0$. The solid
black curve shows the parameters for which the equivalent spin evolution reverses ($dy/dt=0$), for the case of Q=30 and no
atmospheric tides (i.e. the green curve in Figure~\ref{XY}). If we include our atmospheric tide model, we get the solid
red curve for $dy/dt=0$. With the same atmospheric tide but a larger Q=100, the sign reversal locus now shows multiple
branches (the solid blue curves) corresponding to the appearance of new, asynchronous, spin equilibria. The vertical
dot-dashed line shows the edge of the dynamically unstable region.
\label{Eq}}
\end{figure}

The addition of atmospheric tides to the equations changes the spin dynamics of the planet, because the atmospheric
tides are of the opposite sign than the corresponding gravitational tide, and allow for the possibility of asynchronous
equilibria \citep{CL01,Leconte15,CCL15,ALMC17}. Figure~\ref{Eq} shows the consequence of introducing atmospheric tides
into the dynamics of the planet+moon system. Once again, we choose masses appropriate to the Terrestrial system, but
all curves are calculated at $a=0.7$AU.

The dotted curve in Figure~\ref{Eq} shows the locus for $dx/dt=0$, i.e. the change in orbital evolution from inspiral
to outspiral. The solid curve shows the locus of $dy/dt=0$ for the same equivalent case shown in Figure~\ref{XY} -- Q=30
and no atmospheric tides included. The red curve shows the consequence of adding our atmospheric tide model to
the system. Although there is no dramatic topological difference, the spin reversal locus is moved to larger x, and this
can have important consequences for the system evolution.

\begin{figure}
\includegraphics[height=5cm, width=5cm, angle=0, scale=1.75]{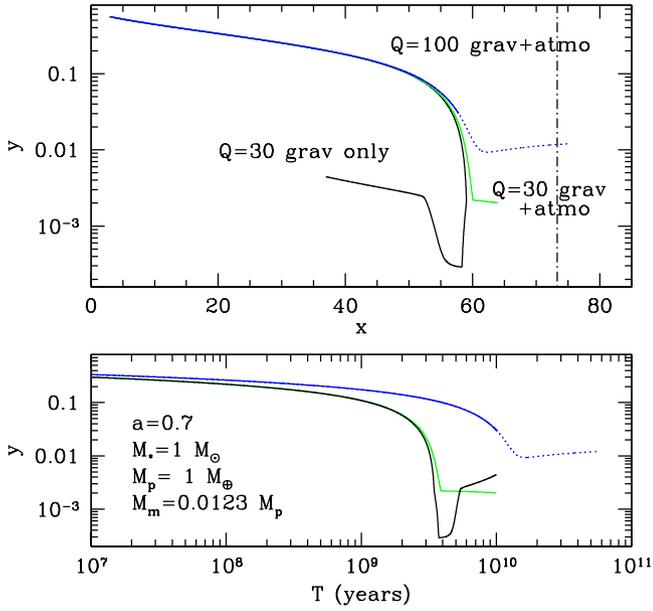}
\caption{The black curve shows the same evolution as the green curve in Figure~\ref{XY}, with Q=30 and no atmospheric tides. In the lower panel
we show the temporal evolution of the spin in each case. The green curve in this figure shows the evolution if we keep Q=30 but include the
atmospheric tide. We see that the character of the evolution is fundamentally changed, as the moon continues to spiral out once atmospheric tides
become important. The cyan curve shows the case for which Q=100 and atmospheric tides are included. The solid curve extends to $10^{10}$ years, and
the dotted curve extends beyond this. We see that the moon will eventually spiral out to escape, driven by the new, asynchronous equilibrium.
\label{XY1}}
\end{figure}

An even more dramatic consequence occurs if we increase Q to Q=100 (weakening the gravitational tide and therefore enhancing the importance
of the atmospheric tide). In this case, the new asynchronous equilibria observed in \cite{Leconte15} -- from which we derived our atmospheric tide model -- appear, and are shown by the blue curves. Of particular interest is the branch that appears above the synchronous curve. Since the evolutionary
curves start at large y and move down as the moon spirals out, they will encounter thi sign reversal first, before the reversal in the direction
of the orbital motion. As a consequence, the injection of angular momentum into the system by atmospheric tides will refresh the outward orbital
evolution. This is shown in Figure~\ref{XY1}, which shows the temporal evolution in each case for the initial conditions used in Figure~\ref{XY}.

The inclusion of the atmospheric tide does indeed refresh the outwards orbital evolution. The action of the atmospheric tide is to weaken the influence
of the tidal forces that spin the planet down, and so this means there is more angular momentum available to drive the planet outwards. This effect
becomes stronger if the gravitational tide is weaker (Q is larger) but the evolution is slower as well because the timescale depends on Q.  This
is explored further in Appendix~\ref{TidalModel}.

It is important to note that, despite the increased complexity, there are still no globally stable equilibria within the Hill sphere. Even in cases where
$dy/dt=0$ in Figure~\ref{Eq}, there are no intersections with the $dx/dt=0$ curve, so that the orbital evolution of the moon will always pull the system out of (or along) the spin equilibria.

\subsection{The influence of tidal strength}

\begin{figure}
\includegraphics[height=5cm, width=5cm, angle=0, scale=1.75]{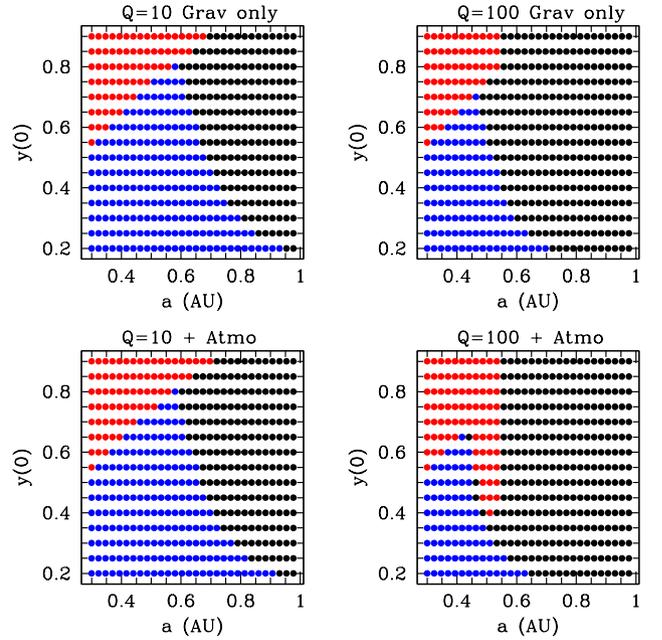}
\caption{The black points indicate systems where the moon survives for 10~Gyr. The red points indicate systems where the moon spirals out to the
point of dynamical instability, while the blue points indicate those systems where the moon ultimately spirals in and is tidally disrupted.
\label{Pans}}
\end{figure}

The uncertainty in the nature and strength of the tidal forces at play means that there is some uncertainty in the exact outcomes in certain parts
of parameter space. To understand the plausible range of variation, we have calculated the evolution of an Earth--Moon analogue system (in terms of mass) for a range of semi-major axis and spin, at both strong (Q=10) and weak (Q=100) dissipation levels, both with and without atmospheric torques parameterised
in the models of \cite{Leconte15}. Figure~\ref{Pans} shows the resulting distribution of outcomes.

For high enough initial spin, there is sufficient angular momentum for moons to become dynamically unstable for semi-major axes $<0.7$AU (with
some dependance on the tidal parameters). For lower values of the spin, the sapping of planetary spin by tides causes inspiral and merger.
Inclusion of the atmospheric tide increases the range over which escape is possible, although the effects of atmospheric tides is weaker if
gravitational tides are stronger. Furthermore, the variation in the strength of atmospheric tide due to atmospheric pressure (as parameterised
in \cite{Leconte15}) yield small shifts in the boundaries between outcomes, but do not introduce substantial qualitative changes.

\subsection{Moon/Planet Mass ratio}
\label{MassRatio}

\begin{figure}
\includegraphics[height=5cm, width=5cm, angle=0, scale=1.75]{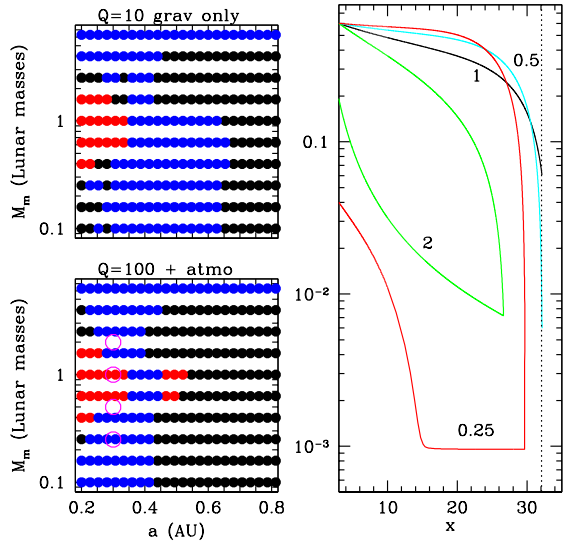}
\caption{The upper left panel shows the eventual outcome for moons of mass $M_m$, orbiting a $1 M_{\oplus}$ planet around a
 $1 M_{\odot}$ star, at the stated semi-major axis. In all cases, the initial planetary spin $y(0)=0.6$. As before, red represents
dynamical instability, blue indicates inspiral and merger, and black represents a moon that survives for 10~Gyr. In this panel,
only gravitational tides are included and are parameterised by Q=10. In the lower left panel, we show equivalent models except
that the gravitational tides are parameterised by $Q=100$ and atmospheric tides are included. The four open magenta circles represent
four models whose evolution is shown in the right hand panel, labelled by the moon mass in Lunar masses. In two cases, the moons reach dynamical instability and in two others,
they result in inspiral and merger.
\label{ascan}}
\end{figure}

The angular momentum balance in a moon--planet system is intimately tied to the mass ratio between the two bodies. 
Figure~\ref{ascan} shows the effect of varying the moon mass while keeping the planet mass fixed. We show two representative
cases, corresponding to the upper left and lower right panels in Figure~\ref{Pans}. These represent the two extermes of
tidal interaction. The case of Q=10 and no atmospheric tides shows systems with efficient tidal interactions, while the case
of Q=100 and atmospheric tides included represent a less efficient tidal evolution. In this case, all initial spins were set
to $y(0)=0.6$, so we probe how the outcomes vary as a function of semi-major axis and moon/planet mass ratio.

We see that dynamical instability occurs only for moon/planet mass ratios similar to those of the terrestrial system. In the case
of very massive moons, this is easy to understand, since the initial angular momentum budget in the spin of the planet is fixed
for all of these models. There is simply not enough angular momentum to drive a massive moon out to Hill sphere distances, and
the planet eventually reaches synchronism and drags the moon inwards. Less intuitively, we find that moons that are too small
also move inwards. This results from the fact that small mass moons are very sensitive to changes in the angular momentum budget
due to planetary spin-down. As we see from the red curve in the right panel of Figure~\ref{ascan}, this evolution initially begins
at the shallowest slope of all, but eventually dips down faster as the stellar tides reduce the planetary spin. We see that the moon mass
is too small to halt the evolution at the synchronous curve and the spin drops down all the way to the stellar synchronism line,
only rising again when the moon spirals in close enough to make its tidal influence significant.

\subsection{Planetary Mass Range}

The potential observability of the catastrophic destruction of a moon will be enhanced if the moon is larger. Thus, we may ask
which combination of plausible parameters yields the largest moons that can escape. We have already seen that we cannot increase
the mass of the moon too much, relative to the planet, before the angular momentum required to reach the instability limit becomes
prohibitive (\S~\ref{MassRatio}). However, we can try to scale up the mass of the planet too.

Although the known planets span several orders of magnitude in mass, our interest here is primarily in rocky planets.
 The moons of
gas giant or Neptune class planets -- those with substantial gaseous envelopes -- may still experience collisions with returning escaped
moons, but the impact of a rocky body into a gaseous atmosphere is likely to excavate little debris that will later be observed as a
dust population. Instead, we might expect a brief optical flash from the impact, but the bulk of the impact energy is expected to be
absorbed by the planet.

\begin{figure}
\includegraphics[height=5cm, width=5cm, angle=0, scale=1.75]{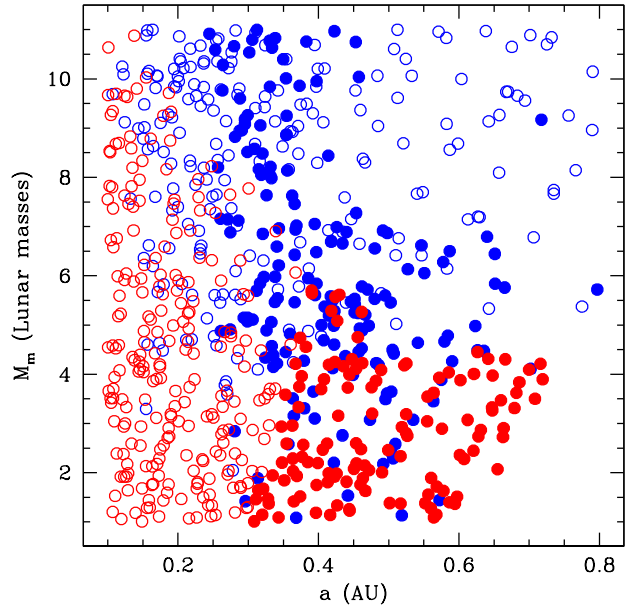}
\caption{The red points show moons that eventually evolved to the point of dynamical instability, while the blue points evolved to
the point of inspiral and merger. All systems orbit a planet of mass $M_p=10 M_{\oplus}$ and $R_p=1.78 R_{\oplus}$, that is hosted
by a 1 $M_{\odot}$ star. Filled circles reach their endpoints at ages $>0.1$~Gyr, while open circles end at earlier ages. The density
of points gets lower at larger separations, as many of the moons at these distances survive for a Hubble time and are therefore not 
plotted here.
\label{Full}}
\end{figure}

It is therefore only for rocky bodies that we expect a significant excavation of debris. The wide variety of exoplanet properties doesn't
lend itself to a clear boundary between rocky and non-rocky bodies, with the observed mass-radius distribution suggesting a continuum
of Hydrogen mass fractions ranging from negligible to dominant. A significant Hydrogen envelope contribution will also affect the radius,
which is the strongest parameter dependance in equation~(\ref{EscapeTime}). Inspection of the
mass--radius relation for known extrasolar planet population \citep[e.g][]{WM14,MCG19,SES19},
restricted to planets with well measured radii and masses, suggests rocky planets can extend up to masses $\sim 10 M_{\oplus}$ and
radii $\sim 1.78 R_{\oplus}$.

Thus, we wish to examine the outcomes for moons orbiting massive rocky planets. Figure~\ref{Full} shows the outcomes for a series of
moon systems around a planet with $M_p=10 M_{\oplus}$ and $R_p=1.78 R_{\oplus}$, assuming $Q=100$ and including the atmospheric tide
model. We allow the semi-major axis to vary between 0.2--0.8~AU, the moon mass to vary between 1 and 10 Lunar masses, and the initial
spin $y(0)$ from 0.2--0.9. Figure~\ref{Full} shows all cases which result in dynamical instability (red) and in merger (blue)

These results demonstrate that we can get dynamical instability for moons up to ten times the mass of the moon, although these require
close-in planets and the moons end their evolution quite quickly. If we also require that the dynamical instability occur at late
times ($> 100$~Myr), then the maximum moon mass is more like $\sim 0.07 M_{\oplus}$.

\subsection{Stellar Mass range}

\begin{figure}
\includegraphics[height=5cm, width=5cm, angle=0, scale=1.75]{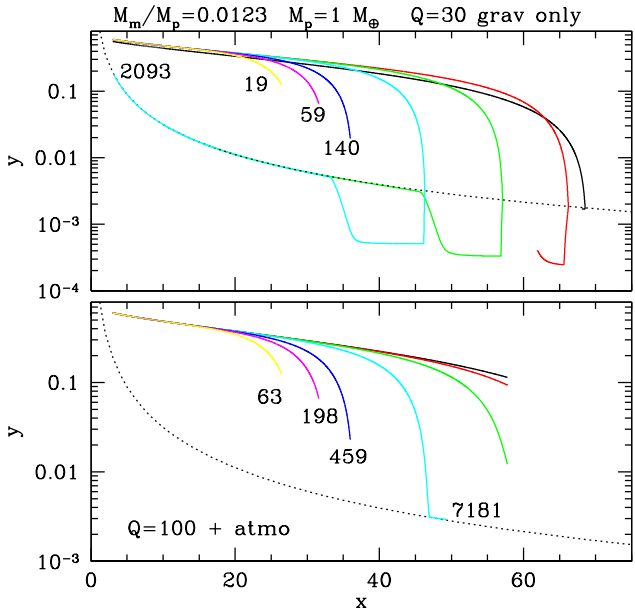}
\caption{The upper panel shows the evolution of an Earth--moon system located around stars of various mass, with semi-major
axes chosen so as to produce the same level of irradiation as the Solar case.
The colours correspond to semi-major axes if 1.0 (black), 0.9 (red), 0.8 (green), 0.7 (cyan), 0.6 (blue), 0.5 (magenta) and
0.4~AU (yellow). The dotted line indicates a synchronous moon orbit. For 0.7--1 $M_{\odot}$, the moon survives to the present day,
while it spirals in and merges at 0.6~AU. Interior to this the moon reaches the point of dynamical instability. The system
age at merger/instability is listed in Myr. The lower panel is equivalent except that the planetary tidal model now uses Q=100 and incorporates atmospheric tides.
\label{MHZ}}
\end{figure}

The mass of the host star also plays an important role in setting both the Hill radius and the strength of the tide acting on the
star. At fixed semi-major axis, a more massive star exerts a stronger gravitational influence and drives tidal and dynamical
evolution on a faster timescale. However, lower mass stars are also less luminous, and so the semi-major axis at fixed luminosity
(such as would be relevant if one were to track the habitable zone as a function of mass -- \cite{KRK13}) decreases rapidly.
The end result is that an Earth-analogue, located in the habitable zone of stars of lower mass than the Sun, must lie closer to
the star and the tidal forces lead to an acceleration in their evolution \citep{Piro18}. Figure~\ref{MHZ} demonstrates this
for the cases of an approximately Earth-like tidal dissipation (Q=30 and no atmospheric tide) and also for a case where the
atmospheric tide is important (Q=100). In each case, the semi-major axis was chosen to match Earth-like levels of irradiation,
given the host star -- using the models of \cite{KRK13}.

We see that, for the more massive host stars, the tidal forces are weak enough that the moon can survive for 10~Gyr without disruption. 
At the lower end of this mass range, the planet does pass through a phase of synchronism with its orbital period, and the moon begins
to spiral inwards. For host stars $\sim 0.6 M_{\odot}$, this inspiral is able to proceed to disruption and merger within 2~Gyr. At
smaller masses, the moon becomes dynamically unstable before synchronism is reached. This is because the dynamical threshold moves inwards
quite rapidly as the host star mass decreases, because the habitable zone distance is a strong function of mass due to the dependance on
stellar luminosity. The timescale to instability also decreases as the host star mass decreases.

This distinction in timescale is relevant for the observational identification of mergers or collisions between moons and planets.
Planetary systems undergo an extended period of scattering and disruption during the process of clearing out the debris from planetary
assembly, and any signature on timescales $<100$~Myr are likely to be confused with the production of dust during the standard debris
disk clearing stage. However, those systems that take several hundred million years, or more, to complete their evolution will be easily
distinguished, as middle-aged stars are not expected to produce significant amounts of dust.

These late-stage events may also be relevant to the habitability of planets themselves. After all, each system in Figure~\ref{MHZ} was
constructed to have Earth-like gravity and Earth-like temperatures, although changes due to tidal despinning and different host star
spectral energy distributions are certainly relevant.
Nevertheless, on these timescales $>$100~Myr, a planet may have started its biological evolution and taken the first steps towards
the construction of an Earth-analogue biosphere. However, tidally induced mergers, or hyperbolic moon collisions (see next section)
may provide a late-type catastrophic event that could quench any start in terms of biological evolution.
 For planets in the habitable zones of these stars, the presence of a moon may eventually wipe
life out, instead of preserving it.

\section{Final Evolution}
\label{Orbits}

In \S~\ref{Unstable} we have reviewed the tidal evolution of the moons orbiting extrasolar rocky planets. As other
authors have noted, the tidal evolution of such systems proceeds faster if the planet is located closer to the
parent star and many moon systems are not expected to survive to the present day. Some are removed by tidally driven
inspiral and disruption, while others spiral out to the point of dynamical instability, at which point they are expected
to become unbound with respect to the planet. We have also included, for the first time, the effects of atmospheric tides
on the moon evolution. We find that this weakens the tidal evolution in some systems, and can increase the fraction
of systems that become dynamically unstable.

Most prior calculations stop at this point, but, if we wish to understand the potential observability of these systems,
we must consider what happens to the moons after they become dynamically unstable. Although they are no longer bound
to the original host planet, they are still bound to the original host star, and we wish to follow their dynamical
evolution beyond the point of instability.

\subsection{Dynamical Instability}

A moon that reaches the dynamical instability limit will transition to a heliocentric orbit, with a similar semi-major axis
as the original host planet. This guarantees close passages with the host planet in the future and the cumulative effect of the gravitational
scattering that result will drive orbital evolution. For a more massive planet, with a larger semi-major axis, such a process
can eventually eject bodies from the system, or deposit them in the Oort cloud. For a terrestrial class planet, deep in a
stellar potential well, the more likely outcome is a planetary collision, with a high probability that it occurs with the
original host planet.

To examine the details of this dynamical disruption process, we have integrated the orbits of potentially unstable moons
 using the direct integration code outlined in \cite{H19}. We begin the moon with circular, coplanar, orbits about the planet with semi-major
axes in the range 0.5--0.6 Hill radii, for a planet mass $1 M_{\oplus}$, in a 1~AU circular orbit about a Solar mass
star. The moon mass is assumed to be negligible, and the orbit is integrated for 1600~years. At this point, the moons
are unbound and the resulting heliocentric orbital
elements are used to start a direct integration of the heliocentric motion of the test particle
moons using the N-body integrator {\em Mercury} \citep{Merc99}. The orbits are then integrated until the test particles
are lost, mostly by collision with the planet. A similar calculation of this phase of evolution was performed by \cite{SAZ19},
which did find some surviving moons, 
but the orbital integration lasted only for 0.5~Myr, which is far too short to assess the long-term stability. 

\subsection{Moon--Planet Collisions}

\begin{figure}
\includegraphics[height=5cm, width=5cm, angle=0, scale=1.75]{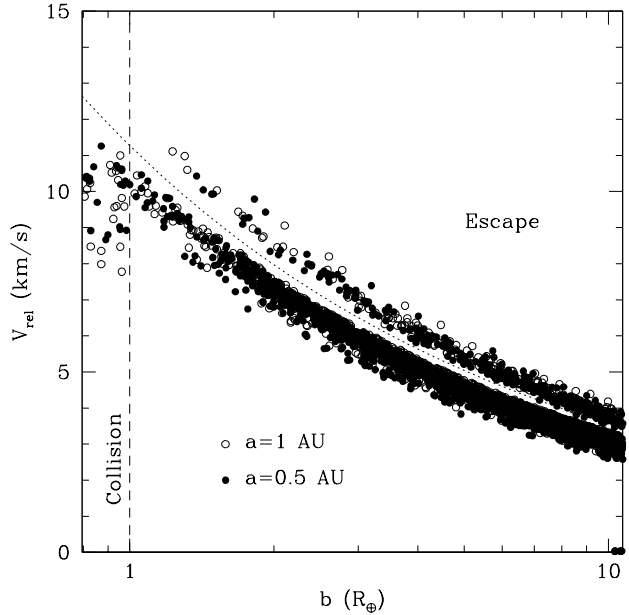}
\caption{Each point shows the relative velocity of the moon and planet during a post-escape close enounter. The ordinate $b$ is the
distance of close approach in each encounter. If $b < 1 R_{\oplus}$, then it represents a true physical collision. The dotted curve
represents the escape velocity from a $1 M_{\oplus}$ mass at distance $b$. Open circles represent a planet with semi-major axis
of 1 AU, while the filled circles show a planet at 0.5~AU. The sample shown here represents 20 realisations of an escaping moon
for each case.
\label{Close1}}
\end{figure}

Indeed, all of our ejected moon candidates eventually collide with the original host planet. Most collide within 10~Myr,
suggesting that the post-instability evolution is a minor contributor to the total age of the system when the collision
occurs. In order to understand the nature of these moon/planet collisions, 
 Figure~\ref{Close1} shows the relative velocity of moon and planet during encounters, as a function of closest
distance in each passage ($b$). We see that most encounters occur at relative velocities below that of the local escape
velocity. This is particularly true for $b<1 R_{\oplus}$, i.e. the terminal close passages that result in a collision.
The amount of mass excavated in a collision will be determined by the energy of the impact, and the fact that this is
not significantly above escape velocity suggests that the mass released is likely to be less than the original moon mass.

This may, in part, be due to the coincidence that the escape velocity from the surface of  Earth is only a factor $\sim 3$ smaller than the
heliocentric orbital velocity (and so comparable to the relative velocity of two bodies with similar semi-major axes). We
therefore also show the case where the planet is closer to the star (0.5~AU) in Figure~\ref{Close1}.  The relative velocity is still dominated by the gravitational
focussing due to the planet (as evidence by the fact that the shape of the distribution follows the dotted curve so closely).
This is ultimately due to the fact that the moons diffuse out of the instability region, rather than getting a significant `kick',
and so their relative velocities with respect to the planet remain quite small, regardless of the semi-major axis. In principle,
it might be possible to increase the relative velocities if we reduce the semi-major axis further, but then the lifetime of the
moon phase becomes comparable to the lifetime of the original planetary assembly phase and so because hard to distinguish observationally.

\subsection{Multiple Planet Systems}

\begin{figure}
\includegraphics[height=5cm, width=5cm, angle=0, scale=1.75]{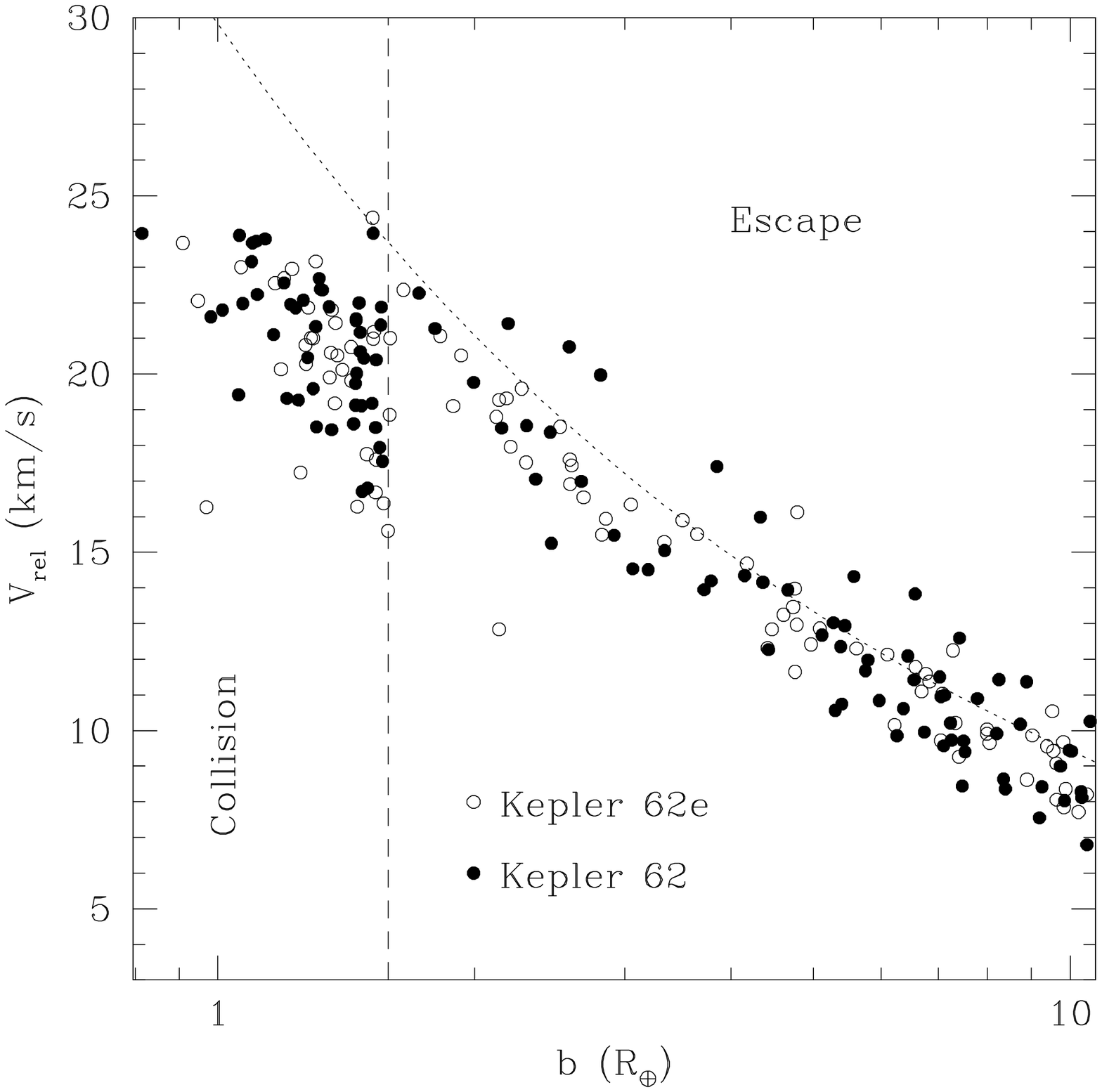}
\caption{Each point shows the relative velocity of the moon and planet during a post-escape close enounter. The ordinate $b$ is the
distance of close approach in each encounter. If $b < 1.7 R_{\oplus}$, then it represents a true physical collision. The dotted curve
represents the escape velocity from a $7 M_{\oplus}$ mass at distance $b$. Open circles represent a planet with the properties
of Kepler-62e orbiting the host star alone. The filled circles show the evolution of the same escaping moons in a system that now
contains three planets -- Kepler-62d, Kepler-62e and Kepler-62f, with the parameters given in the text.
\label{Close2}}
\end{figure}

The parameters of Figure~\ref{Close1} reflect those of an isolated Earth--mass planet on a circular orbit. Many of the real
exoplanet systems contain multiple planets and finite eccentricities. These may, in principle, induce larger relative 
velocities by introducing larger eccentricities. To examine the effects of this, let us consider an analogue of the Kepler-62
system. Although Kepler has discovered many multi-planet systems, most are too compact for this comparison, as they occur at
semi-major axes where the moon evolutionary timescale is too short (see Figure~\ref{Full}). However, Kepler-62e and Kepler-62f
have semi-major axes of 0.427~AU and 0.718~AU respectively. We take Kepler-62e to have mass 7 $M_{\oplus}$ and radius $1.7 R_{\oplus}$.
In Figure~\ref{Close2} we show the close passage velocities of test particle moons released from Kepler-62e. The open circles
represent the case of the single isolated planet, while the filled circles indicate the case where Kepler-62e orbits together
with two additional planets, Kepler-62~d (0.120~AU, 10$M_{\oplus}$ and e=0.1) and Kepler-62~f (0.718~AU, 7$M_{\oplus}$ and e=0.1).
This represents the kind of secular evolution experienced by a compact planetary system.

Figure~\ref{Close2} shows that there is some more spread in velocities in the multi-planet case, but not sufficient to
make a qualitative difference in the behaviour of collisions. The lifetimes of the ejected moons are also shorter in the
multi-planet system, as the precession induced by the additional contributions to the gravitational potential shorten the
intervals between close encounters. 
 Our fundamental conclusion is that almost all moons that are
lost through dynamical instability will end up colliding with their original host planet, at velocities of similar order
of magnitude to that of the escape velocity from the planet.


\subsection{Merger}

In cases where the stellar tides come to dominate the evolution, the planet reverses its migration and starts to
spiral inwards, spinning the planet up. The eventual outcome of this process is a merger or tidal disruption. The
most likely outcome for this process is tidal disruption, because moons generated by impacts tend to be made of
lower density crustal material from the host planet. This disruption may result in the formation of a set of rings \citep{Piro18}
or a set of smaller moons \citep{HM17}. Such an episode is likely to be far less dramatic in terms of dust production than
the high speed impact associated with a collision.

\section{Observability}
\label{EDD}

The essential consequence of the previous sections is that moons formed around rocky planets on scales $\sim 0.4$--0.8~AU
are likely to become dynamically unstable within a Hubble time. When such moons are lost from their parent systems, they
are overwhelmingly likely to collide again with their parent planets, on a much shorter timescale than their tidal evolution.
Thus, the most likely observational consequence is the event associated with the collision.

Collisions between rocky bodies are expected to produce debris that will collide and grind down to dust, which absorbs and
reprocesses stellar light, leading to an infrared excess. This is a generic feature of young planet forming systems, as
collisions between rocky bodies are integral part of the planet formation process. However, such excesses become increasingly
rare with advancing stellar age, as planetary systems mostly complete their assembly within 100~Myr. Thus, excesses on timescales
of Gyr or greater are an indication of some additional processes causing collisions.

Indeed, there are reports of such signatures amongst a subset of older stars \citep{RSZ08,Melis16,MOS21}, where excesses of near infrared emission 
are reported around stars with estimated ages of 100~Myr or more. In order to estimate the size
of the infrared excess expected from our model, we note that the impact velocities $V_{imp}$ calculated in Figure~\ref{Close1} and \ref{Close2} are of order
$\sim 70\%$--$90\%$ of the escape velocities from the surface of the planet.  Thus, equating incoming kinetic energy of an impactor of mass $M_{imp}$ with the amount
of energy required to unbind an amount of material $\Delta M$ from the planet,
\begin{equation}
\frac{1}{2} M_{imp} V_{imp}^2 \sim \alpha_{bind} \frac{G M_p \Delta M}{R_p}
\end{equation}
we find that
\begin{equation}
\frac{\Delta M}{M_{imp}} \sim \frac{1}{2 \alpha_{bind}} \frac{ V_{imp}^2 R_p}{G M_p} 
\end{equation}
which is of order unity, since $V_{imp} \sim (G M_p/R_p)^{1/2}$ and $\alpha_{bind} \sim 1/2$. Thus, an impactor like our
moon should generate a mass $\sim 0.01 M_{\oplus}$, with possibly even a factor of 10 larger in the case of more massive
planet--moon systems.

If this dust is broken into particles of size $s$, it could, in principle, cover an area
\begin{equation}
 A_{IR} \sim 1.7 \times 10^{28} cm^2 \left( \frac{M_{dust}}{0.01 M_{\oplus}} \right) \left(\frac{\rho}{3 g/cm^3} \right)^{-1}
\left( \frac{s}{10 \mu m} \right)^{-1}
\end{equation}
which could render a star completely opaque out to semi-major axes $\sim 2$AU. However, the generation of small dust requires that the
debris released from the collision grind down in a collisional cascade, so the amount of dust present in the system
at any instant is determined by the rate at which material is ground down.

If the moon debris is initially broken up into pieces with characteristic radius $R \sim 1$km, then we have approximately
\begin{equation}
N_{rocks} \sim 2.9 \times 10^9 \frac{M_{dust}}{0.01 M_{\oplus}} \left( \frac{R}{1 km} \right)^{-3}
\end{equation}
each with a mass $\sim 2 \times 10^{16}g (R/1 km)^3$. 

We can estimate the collision rate of this debris by assuming they are spread around a torus of semi-major axis
comparable to the planetary orbit and thickness  comparable
to the planetary Hill sphere radius. Relative velocities between dust grains
are estimated to be  $\sim 10\%$ the local circular velocity. This yields
a collision rate
\begin{equation}
\Gamma \sim 3.9 \times 10^{-4} {\rm yr^{-1}} \frac{M_{dust}}{0.01 M_{\oplus}} \frac{1 km}{R} \left( \frac{a}{1 AU} \right)^{-3.5}
\left( \frac{M_p}{M_{\oplus}} \right)^{-2/3} \left( \frac{M_*}{1 M_{\odot}} \right)^{7/6}
\end{equation}
Note that the collision rate increases as $R$ goes down, so we can assume that the collisional cascade proceeds rapidly to dust
size scales. This gives us a rate at which the collisional cascade generates particles of size $s=10 \mu m$, namely
\begin{equation}
\dot{N} \sim 1.8 \times 10^{30}  yr^{-1} \left(\frac{M_{dust}}{0.01 M_{\oplus}}\right)^2 \frac{1 km}{R} \left( \frac{1 AU} {a} \right)^{3.5}
\left( \frac{M_{\oplus}}{M_p} \right)^{2/3} \left( \frac{10 \mu m}{s} \right)^{3}
\end{equation}

The amount of dust mass present in the system will be regulated by the processes that remove dust from the system. 
For dust of size $s$, in orbit at semi-major axis $a$ around a star of luminosity $L$, the timescale to spiral inwards
due to the Poynting-Robertson effect is
\begin{equation}
T_{PR} \sim 8.4 \times 10^4 {\rm yrs} \left( \frac{a}{1 AU} \right)^2 \frac{s}{10 \mu m} \frac{L_{\odot}}{L}
\end{equation}
The dust can also undergo its own collisional evolution and grind down to the point where it is removed by radiation pressure.
Following the arguments of Zuckerman \& Song \citep{ZS12}, we derive an even shorter lifetime
\begin{equation}
T_{coll} \sim \frac{1.6 \times 10^{-3}}{\tau} {\rm yrs} \left( \frac{M_p}{M_{\oplus}} \right)^{1/3} \left( \frac{M_*}{M_{\odot}} \right)^{-5/6}
\left( \frac{a}{1 AU} \right)^{3/2}
\end{equation}
where $\tau$ is the total  optical depth of the dust cloud around the star. This estimate follows that of \cite{ZS12} except that the width
of the torus of dust is here assumed to be the Hill radius of the planet, since the dust has a localised origin instead of being
generated by a primordial belt of planetesimals.

If we multiply the dust generation rate $\dot{N}$ by the lifetime $T_{coll}$, we find that the overall optical depth of the dust
population generated by this collision is
\begin{eqnarray}
\tau &\sim & 2\times 10^{-3} \frac{M_{dust}}{0.01 M_{\oplus}}
\left( \frac{R}{1 km} \right)^{-1/2} \left( \frac{a}{1 AU} \right)^{-2} \nonumber \\
&& \left( \frac{M_p}{1 M_{\oplus}} \right)^{-1/6} 
 \left( \frac{M_*}{M_{\odot}} \right)^{1/6}
\left( \frac{s}{10 \mu m} \right)^{-1/2}.
\end{eqnarray}

The expected lifetime of the infrared excess can be obtained by estimating the time required to consume the reservoir of N
objects created by the collision, assuming the collision rate $\Gamma$. This leads to a lifetime
\begin{equation}
T_{IR} \sim 2.4 \times 10^3 {\rm yrs} \left(\frac{M_{dust}}{0.01 M_{\oplus}}\right)^{-1} \frac{R}{1 km} \left( \frac{a}{1 AU} \right)^{3.5}
\left( \frac{M_p}{M_{\oplus}} \right)^{2/3} \left( \frac{M_*}{1 M_{\odot}} \right)^{-7/6} \label{AgeIR}
\end{equation} 

The location of the dust will also determine the temperature of reradiated emission, by virtue of the equibrium between absorbed
and emitted flux
\begin{equation}
T_{dust} \sim 280 K \frac{T_{eff}}{5780 K} \left( \frac{R_*}{R_{\odot}} \right)^{1/2} \left( \frac{a}{1 AU} \right)^{-1/2}
\end{equation}
For dust on scales $\sim$0.4--0.8~AU around Sun-like stars, temperatures span 300--450~K, falling into the category of `Warm dust'.

The age of the system when the moon becomes unstable is a strong function of planetary semi-major axis, and can range from
$<10$Myr to $>1$Gyr (see Figure~\ref{Full} for example). As such, those systems for which the tidal evolution is sufficiently rapid
will likely be indistinguishable from a more traditional scenario in which the dust is released by collisions during the actual
assembly of the planetary system. However, the age distribution of systems with unbound moon collisions is expected to have a 
longer tail to Gyr-level ages. To illustrate this, we can estimate the frequency of rocky planetary systems as a function of
semi-major axis as $df/da \sim $constant, based on the empirical observation that the distribution with period is $df/dP \sim P^{-1/3}$\citep{Y11}.
Given the strong dependance of $T_{esc}$ on $a$ (equation~\ref{EscapeTime}), we anticipate that the distribution of systems with
escape time $T$ should scale as $f(T) \propto T^{-11/13}$, which falls off much slower than the usual exponential that is fitted
to the lifetimes of young star infrared excesses. 

The features of warm temperatures, large optical depths and potentially ages up to several Gyr make this a natural scenario
for the origin of dust observed around  older Solar-type stars. The frequency of occurrence of such excesses drops with system age, as
expected for a phenomenon associated with the final accumulation of rocky planetary systems. As such, we expect most systems with
extreme debris disks to be younger than 100~Myr. Nevertheless, there are a handful of objects which are demonstrably much older,
 such as BD+20~307 \citep{RSZ08,WBSZ11}, $\eta$~Corvi \citep{LWC12} or TYC~8830~410~1 \citep{MOS21}. To estimate the frequency
of such objects, we note that the survey of \cite{Moor21} found an upper limit of 6 candidates out of 78650 sources identified as F5--G9 sources with
ages $>100$~Myr. This indicates an occurrence rate $\sim 8 \times 10^{-5}$, or a characteristic lifetime $\sim 8 \times 10^3$~years or less.
This matches well with the lifetime estimated in equation~(\ref{AgeIR}).

 These late-time excesses have spurred suggestions of late time asteroid belt collisions, comet disruptions or analogues of the Late Heavy Bombardment.
Here we suggest that the tidal evolution, dynamical instability and subsequent collision of the unbound moon offers an
alternative scenario that matches the temperatures and optical depths observed for the EDD sample, while also offering
a natural explanation of the timescale -- a feature lacking from the other scenarios.

\section{Discussion}

Satellite systems can teach us much about the origins of planets, but are usually well below the detection thresholds of present day astronomical instrumentation. 
Direct observations are now beginning to probe the most massive satellites orbiting extrasolar giant planets, but direct observation of the moons of terrestrial
class exoplanets will probably remain impossible for some time. Nevertheless, indirect observation of moons or the consequences of their presence, may provide
some information on this topic. In this spirit we have examined the evolution of moons of rocky exoplanets, subject to both gravitational and atmospheric tides,
and also followed this evolution to the ultimate fate of the moon.

\begin{enumerate}
\item 
As has been shown before,  over much of the extant parameter space, the tidal evolution of moons brings them to the point of orbital instability,
with the moon becoming unbound from the original host planet and entering a heliocentric orbit about the host star. We have shown that the
inclusion of atmospheric tides aids this process by opposing the gravitational torque from the star and thereby preserving more of the original
angular momentum to drive the outward orbital evolution of the moon. \\
\item 
We have shown that the overwhelmingly most likely outcome of this orbital instability is that the moon will return, in short order, to collide with
the original host planet on a near-parabolic orbit. Such impacts are expected to generate a cloud of dust that can lead to an infrared excess. For certain
parts of the parameter space, these excesses may occur for ages of several Gyr -- much later than expected for the collisions associated with rocky planet
assembly -- and so should be identifiable on the basis of their unusual time of occurrence. \\
\item Indeed, there is an emerging class of objects, entitled Extreme Debris Disks (EDD) which exhibit all the signatures we associate with these late time, moon--planet
collisions. They occur around stars of intermediate to old age, they show dust temperatures which place them within an AU of the host star (so that tides will 
operate fast enough) and show optical depths $\tau \sim 10^{-4}$--$10^{-3}$, consistent with the amount of dust expected from the collision of a moon-sized object.
Other scenarios have been proposed to generate the collisions required to explain EDD \citep{Wyatt08,Melis16,MOS21}, but all require some level of coincidence to trigger the phenomenon -- the
onset of a late time dynamical instability (which would require a special set of initial conditions) or the injection of a comet from an external debris belt (which
would require another perturber to operate -- \cite[e.g.][]{WSG07}). The moon-based scenario has the distinct advantage of inevitability, in that a moon, once formed, will evolve inexorably
according to the angular momentum budget of the system and the strength of the tides operating on it. \\
\item If  EDD are the result of collisions of tidally unbound moons with planets, then they attest to the occurrence of moons in extrasolar systems
containing rocky planets. Although such moons are a natural occurrence in formation scenarios that include a phase of late stage in situ assembly, they are not
guaranteed in scenarios where the planets migrate inwards from more distant locations. This migration is expected to occur during the epoch when the protoplanetary
gas disk remains and the torques exerted by the gas disk may render such orbits unstable. The existence of moons in extrasolar planet systems would argue that there
was a late stage of planetary collisions in these systems, either due to a genuine episode of in situ assembly \citep{HM12,HM13,CL13} or as the result of the dynamical instability of compact
resonant chains \citep{IOR17}. \\
\item The presence of the moon in the Terrestrial system is believed to aid the habitability of the Earth by stabilising it against obliquity variations, although some authors suggest the influence is more neutral. However, in those systems which form a moon that later goes dynamically unstable, the formation of moons may point in the completely opposite direction -- this may ultimately prove to be harmful for life. The return of an unbound moon on timescales $\sim $Gyr means that the planet would experience a catastrophic impact. Such a collision  would likely sterilise a planet that had begun to form life in the 
time it took for the moon to leave and return. Whether life could then start again after this collision is uncertain, especially if such late time collisions remove much of the water from a potentially habitable planet.
\end{enumerate}

{\bf Data availability}: The data underlying this article will be shared on reasonable request to the corresponding author.

 The author appreciates the constructive comments of the referee.
This research has made use of the NASA Exoplanet Archive, which is operated by the California Institute of Technology, under contract with the National Aeronautics and Space Administration under the Exoplanet Exploration Program. This research has made use of NASA’s Astrophysics Data System, which is 
operated by the Smithsonian Astrophysical Observatory under NASA Cooperative Agreement 80NSSC21M0056.

	\bibliographystyle{mnras}
	\bibliography{refs}

\appendix

\section{Atmospheric Tide Model}
\label{TidalModel}

The thermal inertia of a planetary atmosphere means that the mass redistribution of the atmosphere, in response to the heating from the central star, lags the true diurnal cycle. This means that the torque, applied by the atmosphere to the planetary interior through boundary layer frictional forces, also lags and contributes an asynchronous contribution to the total torque \citep{GS69,CL01,Leconte15,CCL15,ALMC17} . Although such effects are minor for Earth-like planets, they become progressively more important for planets with high levels of irradiation (and are thought to be the determining factor in the retrograde spin of Venus).

To model atmospheric tides over a broad range of irradiation, we adopt the parameterised model from \cite{Leconte15}. They performed a series of global circulation models to quantify the strength of the atmospheric torque as a function of surface pressure $p_s$ (a proxy for atmospheric mass), composition, and incident stellar flux $F_{in}$. To these simulations they fit an analytic model which we adopt as function $ b_a(n_p,\Omega_p)$ in equation~(\ref{eqn:AtmoT}), namely
\begin{equation}
b_a(n_p,\Omega_p) = \frac{(\Omega_p-n_p)/\omega_0}{1 + \left[(\Omega_p - n_p)/\omega_0\right]^2}
\end{equation}
which is parameterised by a characteristic frequency response $\omega_0$ that is a function of $p_s$, $F_{in}$ and composition. The strength of the atmospheric response is also quantified by an amplitude function $q_a$ (see equation~(\ref{eqn:AtmoT})), which is also a function of $p_s$, $F_{in}$ and composition.

\begin{figure}
\includegraphics[height=5cm, width=5cm, angle=0, scale=1.75]{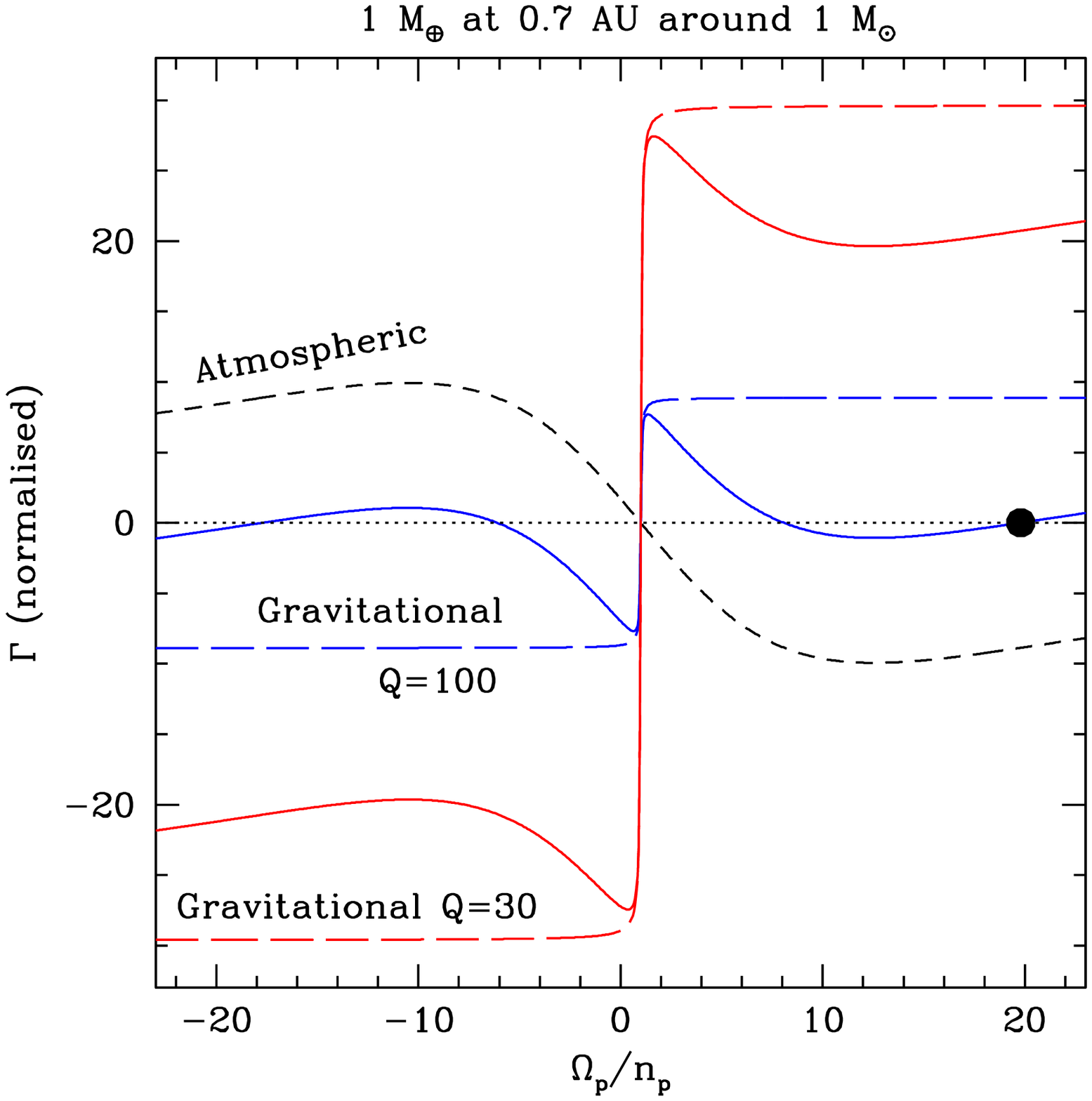}
\caption{ The long-dashed line shows the solid body torque for a $1 M_{\oplus}$ planet, orbiting in a circular orbit at 0.7~AU from a 1 $M_{\odot}$ star.
The red curve is for $Q=30$ and the blue curve is for $Q=100$.
The short-dashed line is the atmospheric torque operating on the same planet, based on the default model described in the text. The solid line represents the net torque operating on the planet, with the blue and red curves corresponding to the sum of the corresponding solid body torque with the same atmospheric torque. Where these curves cross zero (the dotted line), we have a potential spin equilibrium. The filled solid point indicates the first equilibrium the system would encounter as the planet spins down due to the outspiral of a moon.
\label{AppendFig}}
\end{figure}

\cite{Leconte15} provide a table of $q_a$ and $\omega_0$ for a series of different simulations. Our focus is on planets similar to Earth but closer to the Sun, so we adopt the values for $p_s$=1~bar, Earth-like composition, and $F_{in}=1366 W/m^2$. This yields $q_0=1180$~Pa and $2\pi/\omega_0 = 32$~days.
The atmospheric torque also operates in the opposite direction to the direct gravitational solid-body torque, opening the possibility of an asynchronous equilibrium, as discussed by \cite{Leconte15}. Figure~\ref{AppendFig} demonstrates how this arises.

The short dashed curve represents the atmospheric torque model above. The long dashed line indicates the equivalent solid body torque, 
 using $\bar{b}_g = arctan(26.8 (y - 1))$ -- the numerical value of equation~(\ref{barbp}) in this case. The red curve indicates the case
of $Q=30$ and blue indicates the case of $Q=100$. The solid curves indicate the corresponding net torques in the two cases. The parameter
choices for Figure~\ref{AppendFig} are chosen to represent the same cases as in Figure~\ref{Eq}.

The solid black curve in Figure~\ref{Eq} corresponds to the red dashed curve in Figure~\ref{AppendFig} -- the case of no atmospheric torque.
The addition of the atmospheric torque to this (resulting in the red solid line) does not change the qualitative nature of the equilibria,
but weakens the spin down torque and allows the moon to spiral out further before the planetary spin reverses (as seen in Figure~\ref{Eq}).
If the solid body torque is weakened (the blue case), then new equilibria appear.

These asynchronous equilibria can have a significant effect on the moon evolution. Shortly after the formation of the moon, the planet is rotating rapidly and will lie off to the right of Figure~\ref{AppendFig}. As the moon spirals out, the planet spins down and moves to the left in the diagram. In the absence of atmospheric tides, this will continue until the planetary spin is synchronous with the orbital frequency, which lies at $\Omega_p/n_p=1$ in this diagram. However, we see that the system shown in Figure~\ref{AppendFig} will encounter an asynchronous equilibrium first (shown as a filled circle). As the moon continues to spiral outwards (acting to move the system to the left of the solid point), the atmospheric tides act to spin the planet up and move it back to the stable equilibrium. Thus, the atmospheric tides serve to pump angular momentum into the system, rejuvenating the outward motion of the moon, as is seen in Figure~\ref{XY1}.

\end{document}